\documentclass[a4paper,11pt]{article}
\pdfoutput=1 % if your are submitting a pdflatex (i.e. if you have
\usepackage{jinstpub}
\usepackage{lineno}
\usepackage{natbib}
\usepackage{hyperref} 		% makes stuff clickable

\usepackage[separate-uncertainty=true, per-mode=symbol]{siunitx}
\usepackage{booktabs}
\usepackage[version=4]{mhchem} 	% chemistry stuff % set version, otherwise it complains

\DeclareSIUnit{\ton}{t}
\DeclareSIUnit{\ppmm}{ppm(m)}

\newcommand{\scarf}{\textsc{Scarf}}
\newcommand{\llama}{\textsc{Llama}}
\newcommand{\idefix}{\textsc{Idefix}}

\title{Scintillation and optical properties of xenon-doped liquid argon}

\author[*]{C. Vogl,}\note[*]{Corresponding author.}
\author{M. Schwarz,}
\author{X. Stribl,}
\author{J. Grie\ss ing,}
\author{P. Krause,}
\author{and S. Sch\"onert}
\affiliation{Chair for Astroparticle Physics, Department of Physics, Technical University Munich \\
 James-Franck-Str. 1, 85748 Garching, Germany}

\emailAdd{christoph.vogl@tum.de}

\abstract{Liquid argon (LAr) is a common choice as detection medium in particle physics and rare-event searches. Challenges of LAr scintillation light detection include its short emission wavelength, long scintillation time and short attenuation length. The addition of small amounts of xenon to LAr is known to improve the scintillation and optical properties. We present a characterization campaign on xenon-doped liquid argon (XeDLAr) with target xenon concentrations ranging from 0 to \SI{300}{ppm} by mass encompassing the measurement of the photoelectron yield $Y$, effective triplet lifetime $\tau_3$ and effective attenuation length $\lambda_\mathrm{att}$. The measurements were conducted in the Subterranean Cryogenic ARgon Facility, \scarf, a \SI{1}{\ton} (XeD)LAr test stand in the shallow underground laboratory (UGL) of TU-Munich. These three scintillation and optical parameters were observed simultaneously with a single setup, the \textsc{Legend} Liquid Argon Monitoring Apparatus, \llama. The actual xenon concentrations in the liquid and gaseous phases were determined with the Impurity DEtector For Investigation of Xenon, \idefix, a mass spectrometer setup, and successful doping was confirmed. At the highest dopant concentration we find a doubling of $Y$, a tenfold reduction of $\tau_3$ to $\sim\SI{90}{\ns}$ and a tenfold increase of $\lambda_\mathrm{att}$ to over \SI{6}{\m}.}

\keywords{Noble liquid detectors (scintillation, ionization, double-phase), Scintillators, scintillation and light emission processes (solid, gas and liquid scintillators), Large detector systems for particle and astroparticle physics}

\arxivnumber{} % only if you have one

\proceeding{LIDINE 2021: LIght Detection In Noble Elements}

\begin{document}
\maketitle
\flushbottom

%%%%%%%%%%%%%%%%%%%%%%%%%%%%%%%%%%%%%%%%%%
\section{Introduction}

Liquid argon (LAr) is a common detection medium in low-energy rare-event physics searches. It is employed in direct dark matter scintillation detectors (e.g.\ DEAP-3600 \cite{amaudruzDesignConstructionDEAP36002019}) and time projection chambers (e.g.\ DarkSide-50 \cite{darksidecollaborationLowMassDarkMatter2018}), as well as instrumented shielding medium in the neutrinoless double beta decay experiment \textsc{Gerda} \cite{gerdacollaborationFinalResultsGERDA2020},  the upcoming \textsc{Legend-200} experiment and \textsc{Legend-1000}  \cite{ legendcollaborationLargeEnrichedGermanium2017}. LAr features a high stopping power for ionizing radiation and a high light yield \cite{dokeAbsoluteScintillationYields2002} at moderate cost. Its scintillation time structure consist of a fast emission component originating in the singlet state excimer decay with a lifetime of several nanoseconds \cite{adhikariLiquidargonScintillationPulseshape2020, hitachiEffectIonizationDensity1983, peifferPulseShapeAnalysis2008a, carvalhoLuminescenceDecayCondensed1979, kubotaEvidenceTripletState1978} and a slow component created by the decay of triplet state excimers. The reported values of the slow component decay time range from \SI{1.3}{\micro \s} to \SI{1.6}{\micro \s} \cite{adhikariLiquidargonScintillationPulseshape2020, peifferPulseShapeAnalysis2008a, hitachiEffectIonizationDensity1983, kubotaEvidenceTripletState1978, carvalhoLuminescenceDecayCondensed1979}. The population density of the two states depends on the ionization density \cite{hitachiEffectIonizationDensity1983} and enables, due to the large difference in singlet and triplet lifetime, excellent pulse shape discrimination (PSD) performance between electronic and nuclear recoils \cite{amaudruzMeasurementScintillationTime2016, adhikariPulseshapeDiscriminationLowenergy2021, agnesFirstResultsDarkSide502015, hofmannIonbeamExcitationLiquid2013}.

The long triplet state lifetime requires large acquisition time windows to cover the full photon output. Together with the high specific cosmogenic $^{39}$Ar activity of $\sim\SI{1}{\becquerel \per \kg}$ in atmospheric argon \cite{benettiMeasurementSpecificActivity2007, loosliDatingMethod391983}, this can lead to significant dead times in large LAr scintillation detectors at low energies. The scintillation light of LAr is emitted at a low wavelength of \SI{127}{\nm} \cite{heindlScintillationLiquidArgon2010, hofmannIonbeamExcitationLiquid2013}, which is difficult to detect with high efficiency. Finally, the effective attenuation length\footnote{The effective attenuation length $\lambda_\mathrm{att}$ is defined as the wavelength-dependent attenuation length of LAr integrated over LAr's scintillation region.} $\lambda_\mathrm{att}$, is approximately only one meter \cite{ishidaAttenuationLengthMeasurements1997, neumeierAttenuationVacuumUltraviolet2012, neumeierAttenuationMeasurementsVacuum2015}, limiting the maximum size of scintillation detectors.

Adding small amounts of xenon to LAr overcomes these limitations. Xenon-doped liquid argon (XeDLAr) features a longer emission wavelength, higher photoelectron yield $Y$, shorter effective triplet lifetime $\tau_3$ and longer effective attenuation length $\lambda_\mathrm{att}$ than pure LAr \cite{ishidaAttenuationLengthMeasurements1997, neumeierIntenseVacuumUltraviolet2015, akimovFastComponentReemission2019, peifferPulseShapeAnalysis2008a, kubotaSuppressionSlowComponent1993}. The increase in photoelectron yield is partly due to the increase in photodetection efficiency with increasing wavelength. Furthermore, it is expected that the increase in effective attenuation length is due to reduced Rayleigh scattering at longer wavelengths \cite{ishidaAttenuationLengthMeasurements1997}. The scintillation mechanism of XeDLAr relies on the formation of excited argon dimers and subsequent collision with xenon atoms via \cite{kubotaSuppressionSlowComponent1993},
\begin{equation}
\begin{split}
    \ce{Ar^*2 + Xe &-> Ar + ArXe^*} \\
    \ce{ArXe^* + Xe &-> Ar + Xe^*2}.
\end{split}
\end{equation}
The xenon excimers decay with a lifetime of about \SI{4}{\ns} in the singlet state and \SI{20}{\ns} in the triplet state \cite{hitachiEffectIonizationDensity1983}, emitting \SI{175}{\nm} VUV scintillation light \cite{fujiiHighaccuracyMeasurementEmission2015}.

XeDLAr was the subject of many prior investigations \cite{neumeierIntenseVacuumUltraviolet2015, mcfaddenLargescalePrecisionXenon2021, neumeierAttenuationVacuumUltraviolet2015, himpslProjectiledependentScintillationLiquid2020, kubotaLiquidSolidArgon1982, akimovFastComponentReemission2019, kubotaSuppressionSlowComponent1993, wahlPulseshapeDiscriminationEnergy2014, himpslProjectiledependentScintillationLiquid2020, fieldsKineticModelXenonDoped2020, segretoPropertiesLiquidArgon2020, soto-otonImpactXenonDoping2021} which measured subsets of the relevant scintillation parameters, photoelectron yield $Y$, effective triplet lifetime $\tau_3$ and effective attenuation length $\lambda_\mathrm{att}$ for several xenon concentration. However, a combined measurement of $Y$, $\tau_3$ and $\lambda_\mathrm{att}$ was not yet performed and the actual xenon concentration in the liquid phase was not verified. This is important because, in general, the process of injecting xenon into LAr is challenging due to possible loss of xenon through freezing onto cold surfaces. 

In this paper, we present a characterization campaign of XeDLAr with added xenon concentrations ranging from 0 to \SI{300}{ppm} by weight, where the photoelectron yield $Y$, effective triplet lifetime $\tau_3$ and effective attenuation length $\lambda_\mathrm{att}$ were measured simultaneously in a single setup. The actual xenon concentrations in the liquid and gaseous phase during and after doping were determined with a mass spectrometer system.

%%%%%%%%%%%%%%%%%%%%%%%%%%%%%%%%%%%%%%%%%%
\section{Experiment}

The characterization campaign was performed in \scarf, the Subterranean Cryogenic ARgon Facility, a \SI{1}{\ton} (XeD)LAr test stand in the shallow underground laboratory of TU-Munich. Xenon 5.0 (purity $> \SI{99.999}{\percent}$)  was diluted and mixed with argon 6.0 (purity $> \SI{99.9999}{\percent}$) in the gaseous phase  and subsequently, the mixture was injected into the gaseous volume above the LAr phase  in the cryostat. This procedure was developed in order to prevent xenon from freezing onto cold surfaces, as was observed elsewhere \cite{razExperimentalEvidenceTrapped1970}. In total, 5 injections were performed with target xenon concentrations in the liquid phase of 3, 10, 50, 100 and \SI{300}{ppm} by mass. 

Samples were taken from the gas phase via a valve at the top of the cryostat, and from the liquid phase through a valve at the bottom and analyzed with the Impurity DEtector For Investigation of Xenon, \idefix, a quadrupole mass spectrometer system. A certified gas mixture containing \SI{100}{ppm} by volume xenon in argon was used for calibration. The equilibrium xenon concentrations in the liquid and gaseous phases are not identical due to Henry's law. Freezing of xenon onto gas lines or partial evaporation of more volatile components can disturb the liquid phase sample taking process and prohibit the acquisition of representative samples. We used a pre-vacuum pump to enforce rapid evaporation of XeDLAr when determining the xenon concentration in the liquid phase. Enforced, fast evaporation does not allow for chemical equilibrium and hence is expected to  lead to the generation of representative gaseous samples of cryogenic liquids.

The xenon concentration in the gas phase was monitored via continuous measurements where a flow of gas exited the cryostat, passed \idefix\ and was ejected into the atmosphere.

The scintillation properties were characterized with the \textsc{Legend} Liquid Argon Monitoring Apparatus, \llama\ \cite{schwarzLiquidArgonInstrumentation2021}, a silicon photomultiplier array featuring 16 Hamamatsu VUV4 SiPMs (S13370-6075CN) pointing from different distances at a LAr-filled cavity next to an encapsulated $^{241}$Am source. The $^{241}$Am source emits \SI{60}{\keV} gammas inducing (XeD)LAr scintillation light emission. Three SiPMs are located at the side walls of the cavity, at a distance of $\sim\SI{1}{\cm}$ from the source and allow to determine the photoelectron yield $Y$ without having to take attenuation effects into account. The peripheral SiPMs are located at distances ranging from \SI{15}{\cm}  to  \SI{75}{\cm} in steps of \SI{5}{\cm} and allow to measure the effective attenuation length $\lambda_\mathrm{eff}$. The first peripheral SiPM provides timing information to determine the effective triplet lifetime $\tau_3$.

%%%%%%%%%%%%%%%%%%%%%%%%%%%%%%%%%%%%%%%%%%
\section{Results and discussion}

For target xenon concentrations from \SI{50}{\ppmm} on and higher, measured values of the xenon concentration in the liquid phase are available and shown in \autoref{tab:xenon-concentrations}. At prior injections (0, 3 and \SI{10}{\ppmm} target concentration), the experimental method of the forced evaporation technique was not available. Taking into account losses related to the gas sampling, the measured xenon concentration in the liquid phase agrees quantitatively with the amount of inserted xenon. This proves successful doping for \SI{50}{\ppmm} and above and, since the injection process was identical during the whole campaign, provides strong evidence that the prior additions were also successful.

\begin{table}
    \centering    
    \caption{Target and measured xenon concentrations in the liquid phase obtained by \idefix\ where measurements are available. The actual xenon concentrations are lower than the target values in the first two rows. This is attributed to loss of xenon during continuous measurements with \idefix\ from the gas phase where gas with a high xenon concentration is ejected into the atmosphere. This loss was prevented during the last injection; the measured concentration is correspondingly higher.}
    \label{tab:xenon-concentrations}
    \vspace{2mm}
    \begin{tabular}{ll} \toprule
        Target $c_\mathrm{Xe}$ [\si{\ppmm}] &  Measured $c_\mathrm{Xe}$ [\si{\ppmm}] \\ \midrule
        50      & \num{37.9(79)} \\
        100     & \num{87.8(89)} \\
        300     & \num{360(59)} \\ \bottomrule
    \end{tabular}

\end{table}

\begin{figure}
    \centering
    \includegraphics[scale=0.4]{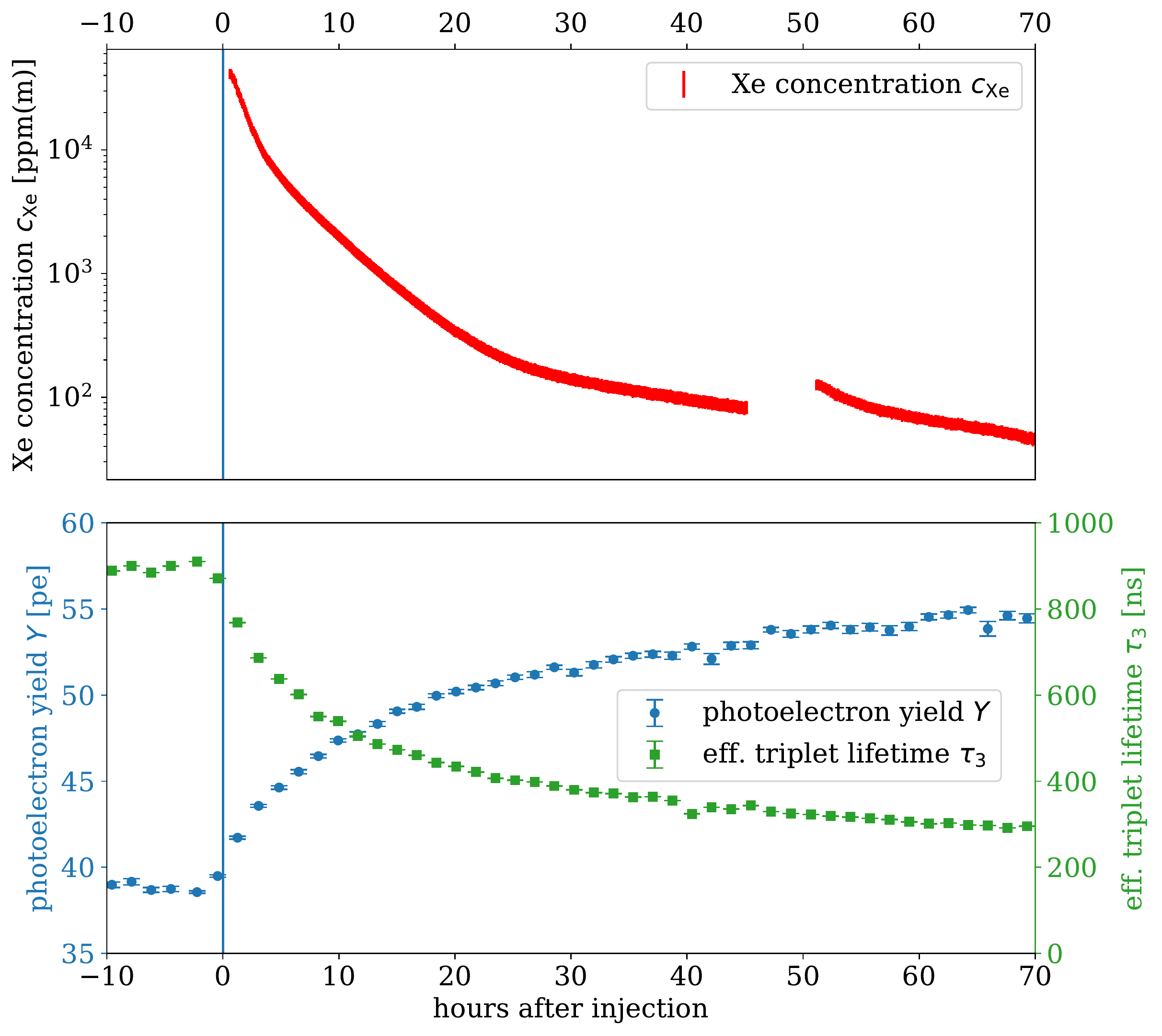}
    \caption{The xenon concentration in the gaseous phase (top), photoelectron yield $Y$ and effective triplet lifetime $\tau_3$ (bottom) after the third argon-xenon mixture injection with a target concentration of \SI{50}{\ppmm} in the liquid phase of the cryostat. The gaseous xenon concentration decreases rapidly, witnessing the transition into the liquid phase. The break in the measurement is due to liquid phase measurements being performed with \idefix\ in that time frame. The scintillation parameters are affected by xenon immediately after injection. $Y$ increases and $\tau_3$ decreases, as expected. The error bars on \llama\ data currently only contain statistic uncertainties. Stabilization occurs after some tens of hours.}
    \label{fig:xenon-mixing-idefix-llama}
\end{figure}

The xenon concentration in the gas phase of \scarf\ just after the third injection (\SI{50}{\ppmm} target concentration) is shown in the top panel of \autoref{fig:xenon-mixing-idefix-llama}. It is rapidly decreasing, witnessing the transition from the gaseous into the liquid phase. At the same time, visible in the bottom panel, the photoelectron yield $Y$ is increasing and the effective triplet lifetime $\tau_3$ decreasing. The parameters stabilize in the course of several tens of hours.

\begin{figure}
    \centering
    \includegraphics[scale=0.4]{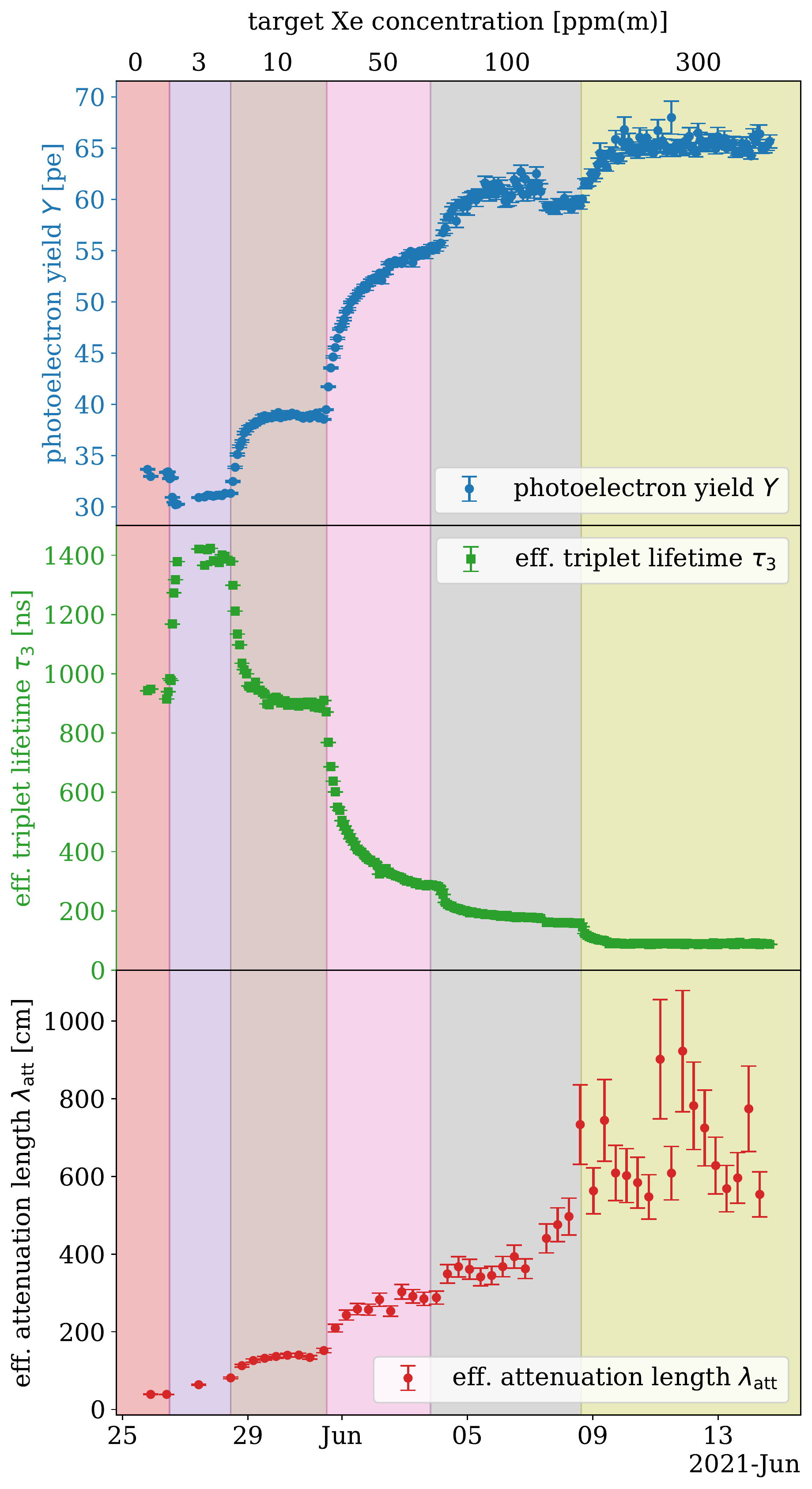}
    \caption{The complete evolution of the scintillation and optical parameters during the XeDLAr characterization campaign. Stable, injection and mixing phases are visible. With each injection the parameters change. $Y$ and $\lambda_\mathrm{att}$ increase, whereas $\tau_3$ decreases. At \SI{3}{\ppmm} a reduction of $Y$ and increase of $\tau_3$ was observed.}
    \label{fig:xenon-mixing-llama-all}
\end{figure}

The complete data set gathered with \llama\ is shown in \autoref{fig:xenon-mixing-llama-all}. $Y$, $\tau_3$ and $\lambda_\mathrm{att}$ are plotted for all xenon concentrations during injection, mixing and stable phases. The initial triplet lifetime of the undoped LAr was \SI{941}{\nano \s} because of the presence of residual impurities. The general trend of parameter development is as expected, the photoelectron yield and effective attenuation length increase and the effective triplet lifetime decreases. At the lowest target xenon concentration, however, an unexpected reduction of $Y$ and increase of $\tau_3$ was observed, which is currently under investigation. The drop at the end of the \SI{100}{\ppmm} target xenon concentration time frame is due to an exchange of pre-amplifiers in the \llama\ electronics. This became necessary because the large observed photoelectron yield exceeded  the dynamic range of the charge sensitive amplifier.

Analyzing the data combined over the time frames of the stable phases, yields the high-statistics reference values shown in \autoref{tab:scintillation-reference-values}. At \SI{300}{\ppmm} the photoelectron yield almost doubles, which is largely due to an increased photodetection efficiency at \SI{175}{\nm} with respect to \SI{127}{\nm} of the used SiPMs \cite{hamamatsuProductFlyerVUVMPPC2017}. The remainder is attributed to recovery of  light lost to the present impurities \cite{acciarriOxygenContaminationLiquid2010, acciarriEffectsNitrogenContamination2010}. The effective triplet lifetime decreases by a factor of $\sim 10$ and the effective attenuation length increases by a factor more than 10 to over \SI{6}{\meter}.

\begin{table}
    \centering
        \caption{Measured scintillation and optical properties of XeDLAr for the investigated target xenon concentrations during stable phases. In general, the photoelectron yield $Y$ and the effective attenuation length $\lambda_\mathrm{att}$ increase, whereas the effective triplet lifetime $\tau_3$ decreases with increasing xenon concentration. The unexpected exception at \SI{3}{\ppmm} is currently under investigation. The quoted uncertainties are of statistical nature.}
    \label{tab:scintillation-reference-values}
    \vspace{2mm}
    \begin{tabular}{llll} \toprule
        Target $c_\mathrm{Xe}$ [\si{\ppmm}] & $Y$ [\si{pe}] & $\tau_3$ [\si{\ns}] & $\lambda_\mathrm{att}$ [\si{\cm}] \\ \midrule
        0   & \num{33.63(5)} & 941  & \num{41.6(9)} \\
        3   & \num{31.10(3)} & 1395 & \num{64(1)}   \\
        10  & \num{39.15(3)} & 883  & \num{139(2)}  \\
        50  & \num{54.51(5)} & 300  & \num{288(8)}  \\
        100 & \num{59.33(6)} & 159  & \num{498(26)} \\
        300 & \num{64.99(4)} & 89   & \num{653(22)} \\ \bottomrule
    \end{tabular}
\end{table}

%%%%%%%%%%%%%%%%%%%%%%%%%%%%%%%%%%%%%%%%%%
\section{Conclusion}

A characterization campaign on the scintillation and optical properties of XeDLAr with target xenon concentrations in the liquid phase from 0 to \SI{300}{ppm} by mass was presented. The photoelectron yield $Y$ and effective attenuation length $\lambda_\mathrm{att}$ increase and the effective triplet lifetime $\tau_3$ decreases with increasing xenon concentration. At \SI{300}{\ppmm}, we observed a doubling of $Y$, a decrease of $\tau_3$ of about a factor of 10 and a more than tenfold increase of $\lambda_\mathrm{att}$. At \SI{3}{\ppmm}, an unexpected decrease of $Y$ and increase of $\tau_3$ was obtained. The actual xenon concentrations in the liquid phase were measured and match the target values, if known losses are taken into account.

Future work will investigate the interesting behavior at \SI{3}{\ppmm} through modeling of the energy transfer process and another characterization run using ultra-pure XeDLAr. The long-term stability of the mixture will be investigated and the absolute photon yield determined.

\acknowledgments

This work has been supported in part by the DFG through the SFB1258.

\bibliography{LIDINE_2021_proceedings_bibtex.bib}

\end{document}